# Density-Functional Theory Study of Hydrogen Induced Platelets (HIPs) in Silicon

Liviu Bîlteanu[1,2] and Jean-Paul Crocombette[1]

[1] Commissariat à l'Energie Atomique et Alternative,
91191 Gif-sur-Yvette Cedex, France.

[2] Laboratoire de Physique des Solides UMR 8502, Université Paris Sud
91405 Orsay Cedex, France.

## ABSTRACT

In this contribution we present the results of Density-Functional Theory (DFT) calculations of platelets as modelled by infinite planar arrangements of hydrogen atoms and vacancies in (100) planes of silicon. From the observation of the relaxed platelet structures and the comparison of their energy with the one of hydrogen molecules dissolved in silicon we were able to evidence several features. A planar arrangement of hydrogen atoms inserted in the middle of Si-Si bonds proves unstable and Si bonds must be broken for the platelet to be stable. In the (100) plane the most stable configuration is the one with two Si-H bonds (a so-called $SiH_2$ structure). It is possible to generate $SiH_3$ structures which are more stable than hydrogen dissolved in Si bulk but less than $SiH_2$ structures but $SiH_1$ or $SiH_4$ sometimes observed in experiments prove unstable.

## INTRODUCTION

Hydrogen induced platelets (HIPs) are bidimensional defects involving 1-2 atomic plans occuring in hydrogen implanted samples of silicon [1] or of some other materials[2]. HIPs have been observed mostly on (100) or (111) plans, being generally oriented within the plans parallel to the implantation surface of the sample [3-4]. Yet, the HIP orientation depends not only on the crystallographic features of the sample but also on the implantation conditions: different implantation conditions can induce different stress gradients into the sample that can influence the orientation of HIPs [5-6]. Several experimental papers have been published related to the occurrence and growth of HIPs (see for example [7] and the references mentioned therein).

Despite the fact that some theoretical works have been published on the structure of HIPs [8-10] there are some open issues related to the HIPs structure and their formation mechanisms. In this paper we present several HIPs models that we have studied by more accurate DFT calculations that have been performed before. We focus on the (100) structures and on their stability in the hydrogen supersaturation conditions. The remaining of this paper is organized in three sections. In the section 2 we shall present the computational details, the methods of the structure generation and the formulas we have used to calculated the formation energies for the generated structures. In the section 3 we present the results we have obtained. Finally, in the last section we shall summarize and draw the conclusions.

## THEORY

All the DFT calculations presented in this paper have been performed with the SIESTA code [11] using norm-conserving pseudopotentials and bases of numerical atomic orbitals. The exchange-correlation functional we have used is the Generalized Gradient Approximation (GGA) [12]. The pseudopotentials were provided with the SIESTA package and the employed basis sets were of double-zeta type. Their radial cut-off values are 7.0 Bohr for Si and 8.0 Bohr for H. The Si basis has been constructed in order to reproduce the properties of the bulk silicon. The lattice parameter is 5.431 Å. The cut-off energy used in the calculations is 150 Ry. The superior limit of the forces after the relaxation of supercells is 0.04 eV/Å. The initial supercell size is 1x1x10, meaning that they are elongated supercells upon the z axis.

Two types of structures have been generated starting from the initial perfect supercell: without removing of Si atoms and with removing of Si atoms. In the first case, H atoms have been added in different concentrations in (100) plans then the supercell has been relaxed. In the second case, where Si atoms are removed, the Si dangling bonds (DBs) are saturated with H atoms and then relaxed. Next $H_2$ molecules have been added inside and outside the HIPs and the newly obtained systems have been relaxed for a second time. Basically the building blocks of the structural models are the most stable structures of hydrogen in silicon: the H atoms in the bond centre (BC) or in the antibond (AB) sites, the $H_2$ molecules in the tetrahedral (T) and hexagonal (Hex) sites, $H_2^*$ diatomic structures (composed of two atoms one in BC site and the other on AB site) and the H-saturated dangling bond.

The total energies calculated using SIESTA have been used to study the stability of HIPs. The stability is quantitatively described by the formation energy. Each HIP can be modeled within a supercell by placing hydrogen atoms (*m*) or by removing Si atoms (*k*) and subsequently saturating the dangling bond by hydrogen atoms. In order to compare the stabilities of different defects, it is better to use a normalized energy; that is an energy divided by the number of elementary defects (vacancies and/or hydrogen atoms) present in the structure. Moreover all the formation energies should be calculated with respect to the same reference. When calculating the formation one can assume or not the pre-existence of vacancies. Depending on these assumptions, the formulas for calculating the formation energy per hydrogen atoms are:

$$E_1^{HIP} = \frac{1}{m}\left[E^{HIP} + \left(\frac{m}{2} - 1 + \frac{k}{80}\right)E^{bulk}\right] - \frac{1}{2}E^{H_2} \quad (1)$$

for the structures created without the pre-existence of vacancies.

$$E_2^{HIP} = \frac{1}{m}\left[E^{HIP} + \left(\frac{m}{2} + k - 1\right)E^{bulk}\right] - \left(\frac{1}{2}E^{H_2} + \frac{k}{m}E^V\right) \quad (2)$$

for the structures created in the context of the preexistence of vacancies.

In these formulas *m* is the number of hydrogen atoms in the structure, *k* is the number of vacancies, $E^{HIP}$ is the energy of the HIP structure, $E^{bulk}$ is the energy of a bulk chunk containing 80 atoms (that is 10 conventional cells), $E^{H_2}$ is the energy of the interstitial molecules and the $E^V$ is the energy of a vacancy.

Molecules are the most stable form of hydrogen dissolved in silicon. Hence if the energies calculated by (1) or (2) are positive then it means that H molecules will prefer not to gather in the specific HIP structure. If the energy is negative then the HIP structure is more stable than the $H_2$ molecules.

In the Discussion section the energies calculated by equation (1) are written first immediately under the respective configurations, while those calculated by equation are written secondly under them.

**DISCUSSION**

In Figure 1 - Figure 3 the first energy is calculated by (1) with respect to $H_{BC}$, while the second energy (in red) is calculated using (2) (hence it is expressed with respect to $H_{BC}$ and Si vacancy). The results are summarized in Table 1 in which we describe briefly the structure and give the energies calculated with both formulas (1) and (2). Except the first structure (Figure 1a), all the structures exhibit negative formation energy per hydrogen atom with respect to both hydrogen atoms in BC only and with respect to hydrogen atoms in BC and Si vacancies.

**Table 1 Summary of the formation energy per hydrogen atom of the HIPs structures in Figures 1-3.**

| Structure description | Structure label | Figure | Energy (eV) no pre-existing vacancies | Energy (eV) pre-existing vacancies |
|---|---|---|---|---|
| **Vacancy-free structures** | | | | |
| Hydrogen atoms in the bond center sites | Si-$H_{BC}$ | Figure 1a | 0.82 | 0.82 |
| Spontaneously created $SiH_2/SiH_2$ surfaces | $SiH_2/SiH_2$ | Figure 1b | -0.19 | -0.19 |
| Surface created by adding one H atoms to $SiH_2/SiH_2$ | $SiH_3/SiH_2$ | Figure 2a | -0.20 | -0.20 |
| Restructured surfaces obtained from $SiH_3/SiH_2$ | $Si_2H_3/SiH_2$ | Figure 2b | 0.02 | 0.02 |
| $SiH_2/SiH_2$ surfaces containing molecules | $SiH_2/H_2/SiH_2$ | Figure 2c | -0.11 | -0.11 |
| **Vacancy-containing structures** | | | | |
| One vacancy layer | $V_1H_4$ | Figure 3a | -0.19 | -1.08 |
| Two vacancy layers | $V_2H_4$ | Figure 3b | -0.19 | -1.98 |

The first structure we have studied is the one built up by an infinite arrangement of $H_{BC}$ atoms in a (100) plan (Figure 1a). The lattice relaxes by 5% in terms of lattice constant. This structure might be very intuitive since $H_{BC}$ is the minimal energy atomic configuration. However, the energy per hydrogen of such structure is 0.82 eV with respect to $H_2$ hence this structure is highly unstable.

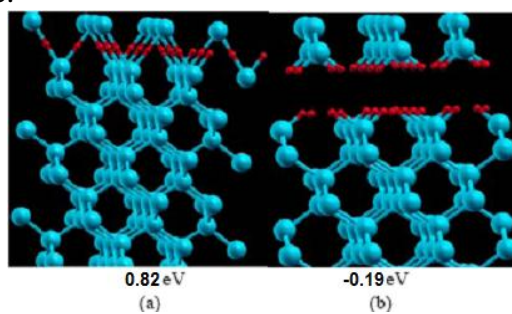

**Figure 1 Hydrogen only structures: (a) Si-$H_{BC}$ structures; (b) $SiH_2/SiH_2$ spontaneously created surfaces.**

Placing $H_2$* complexes or interstitial molecules $H_2$ (in T and Hex sites) in a (100) plan leads to the spontaneous formation of a configuration containing hydrogen dangling bonds (Figure 1b). The energy of such structure is -0.19 eV/H atom showing that is is energetically more stable than $H_2$ molecules: hydrogen prefers to be bounded to DBs in (100) configurations than forming $H_2$ interstitial.

Placing a hydrogen atom in the void space in Figure 1b leads to the increase of the separation space between the two surfaces (Figure 2a). The presence of hydrogen leads to the breaking of a Si – Si bond and the creation of a $SiH_3$ surface. Due to steric constraints half of the $SiH_3$ groups are slightly tilted with respect to the others. The energy of such a structure is -0.20 eV/H atom, quite close to the $SiH_2/SiH_2$ configuration in Figure 1b. However the energy gain from passing from one structure to another is -1.01 eV, in the favour of the $SiH_3/SiH_2$ configuration.

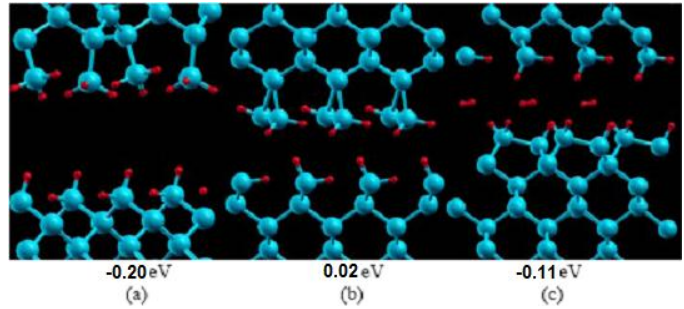

**Figure 2 (a) SiH$_3$/SiH$_2$ surface; (b) Si$_2$H$_3$/SiH$_2$ surface; (c) hydrogen molecules inside the SiH$_2$/SiH$_2$ surface represented in Figure 1b (SiH$_2$/H$_2$/SiH$_2$).**

Removing only one H atom from the superior surface of the SiH$_2$/SiH$_2$ structure (Figure 1b) and then relaxing the newly obtain structure one observes a reorganisation of the Si surface and the creation of a higher energy (0.02 eV) configuration of the type Si2H$_3$/SiH$_2$ (Figure 2b) is obviously ruled out.

The most straightforward configurations containing vacancies are those obtained by removing one layer of Si atoms and saturating the DBs by H atoms. The energy of H atoms within the two configurations in Figures 3a and b is the same (-0.19 eV) as that of the structure obtained without removing any layer of Si atoms (Figure 1b) and the SiH$_2$ groups are separated by the same distance. This means that the structure of the Si-Si interface does not count for the HIP energy. However with respect to an implantation damage system, the V$_2$H$_4$ configuration is more stable (by 0.90 eV) with respect V$_1$H$_4$ configuration.

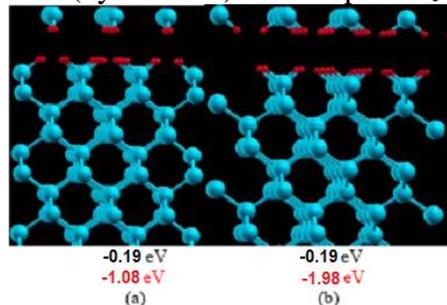

**Figure 3 Vacancy containing HIPs models: (a) V$_1$H$_4$ structure obtained by removing one layer of Si atoms; (b) V$_2$H$_4$ structure obtained by removing 2 layers of Si atoms.**

Adding molecules inside the spontaneously created SiH$_2$/SiH$_2$ structure in Figure 1b leads to a further spacing between the two surfaces (Figure 2c). Structurally one observes that the SiH$_2$ groups are slightly tilted due to steric constraints imposed by the accommodation of molecules. The energy of such a configuration (-0.11 eV/H atom) is greater than the perfect SiH$_2$/SiH$_2$ configuration, so unfavourable with respect to the later one, but still low enough to make this configuration more favourable than the interstitial H$_2$ molecule.

Introducing molecules in the void space of the V$_1$H$_4$ type structure increases the energy calculated by (2) by 0.19 eV. This increase of the energy is due to the fact that hydrogen in H$_2$ is a state less stable than the Si-H bond itself. The trend stays the same when one adds another H$_2$ molecule layer in order to obtain the VH$_2$/2H$_2$ configuration.

Additional hydrogen outside extended structures leads to the destabilisation of HIPs. For example, adding H$_2$ to the SiH$_2$/SiH$_2$ structure in Figure 1b obtained through spontaneous relaxation is leading to an destabilization by an energy increase of 0.10 eV. The SiH$_3$ surface in Figure 2c is stable in the presence of H$_2$ molecules but the energy increase is of 0.03 eV. Hydrogen molecule insertion inside the structure in Figure 3a decreases significantly its energy by 0.26 eV. Adding hydrogen molecules in the void space inside the structure in Figure 3b decrease the energy by 0.08 eV. The presence of hydrogen molecules does not change the hydrogenation state of the surfaces of the initial configuration.

Most of the structures proposed within this paper are lower in energy than the $H_2$ molecules in interstitial sites. This enters in a flagrant contradiction with the results reported by Martsinovich et al.[9] who report that all models have an energy value greater than the $H_2$ molecule in the T site. basis set. We can reasonably hope to have a greater precision than this previous calculations due to larger supercells, finer k-space sampling and a set of suitable and well-test numerical atomic orbitals.

**CONCLUSIONS**

Several HIPs models have been investigated through first principle calculations. These calculations allowed establishing the structural features of the HIPs. The major result in this work is that hydrogen prefers essentially to create surfaces inside the Si bulk. All stable HIP structures involve a surface (formed by $SiH_2$ groups and sometimes by $SiH_3$ groups) . No surface free HIP has been detected and the HIPs involve usually up to two Si layers. The structure which does not contain hydrogenated dangling bonds (Figure 1a) is highly unstable.

In the case of the structures created by the removal of one or two Si layers, the energy gain favours tremendously the structure from which two Si layers have been removed. The enormous energy gain (per H atom) of 1.08 eV and 1.98 eV respectively shows that the HIPs are actually self-building vacancy sinks: both vacancies and hydrogen species are stabilised by joining these structures. By energy considerations, it has been proved, hence that HIPs are self-building potential sinks for Si vacancies and H atoms. When all the dangling bond are saturated inside a HIP, further accumulation of hydrogen leads to the formation of molecules that are contributing to the HIPs growth and further the fracture

Hydrogen insertion inside HIP void space tends to destabilise the HIP which remains more stable than isolated $H_2$ molecules though. Such energy considerations confirm the model of a $H_2$ driven growth of HIP and, further of the fracture. In such a model $H_2$ molecules are created inside the HIPs by systematically capturing all the hydrogen species (atoms or molecules) inside the bulk. The $H_2$ molecules form a gas, whose pressure further increases the size of the HIP in the (100) plan and increases the spacing between the two surfaces of the HIP leading finally to the fracture.

Our results show that the presence of hydrogen atoms and molecules outside the HIP tend to actually destabilise the HIPs. Such structures exhibiting hydrogen outside the void space and surfaces are less stable than the structures containing hydrogen molecules inside. This means that, nearby a HIP, hydrogen tends immediately to be incorporated into that HIP first to saturate any existing dangling bond and afterwards to form molecules inside a HIP.